\documentclass[aps,twocolumn,prl,showpacs]{revtex4}
\usepackage{graphicx}
\begin{document}
\draft
\preprint{1 July 2006}
\title{Spin dynamics in molecular ring nanomagnets:
       Significant effect of acoustic phonons and magnetic anisotropies}
\author{Shoji Yamamoto and Toshiya Hikihara}
\address{Department of Physics, Hokkaido University,
         Sapporo 060-0810, Japan}
\date{1 July 2006}
\begin{abstract}
The nuclear spin-lattice relaxation rate $1/T_1$ is calculated for
magnetic ring clusters by fully diagonalizing their microscopic spin
Hamiltonians.
Whether the nearest-neighbor exchange interaction $J$ is ferromagnetic
or antiferromagnetic, $1/T_1$ versus temperature $T$ in ring nanomagnets
may be peaked at $k_{\rm B}T\simeq |J|$ provided the lifetime
broadening of discrete energy levels is in proportion to $T^3$.
Experimental findings for ferromagnetic and antiferromagnetic
Cu$^{\rm II}$ rings are reproduced with crucial contributions of magnetic
anisotropies as well as acoustic phonons.
\end{abstract}
\maketitle

   In recent years, considerable efforts \cite{G1054} have been devoted to
constructing and investigating magnetic systems of nanoscale dimension
that comprise a controllable number of transition metal ions.
Resonant magnetization tunneling \cite{F3830,T145} in
[Mn$_{12}$O$_{12}$(CH$_3$COO)$_{16}$(H$_2$O)$_4$], now well-known as Mn12,
has stimulated increasing interest in mesoscopic magnetism.
Among such topical molecular nanomagnets are highly symmetrical clusters
of almost planar ring shape \cite{C182,S277}.
They are highly varied, containing four to eighteen metal ions of spin
$\frac{1}{2}$ to $\frac{5}{2}$ and therefore, serve to reveal a
quantum-to-classical crossover between molecular and bulk magnets
\cite{H054409}.

   Nuclear magnetic resonance (NMR) is an effective probe of
low-energy spin dynamics, and the nuclear spin-lattice relaxation time
$T_1$ has been measured for various molecular wheels
\cite{L14341,L1115,L6946,J227,L6839,B134434,P184419,M020405,M034710}.
Interestingly, $1/T_1$ as a function of temperature $T$ under a fixed
field $H$ is commonly peaked at $k_{\rm B}T\simeq |J|$, where $J$ is the
intracluster exchange interaction between neighboring ion spins.
Baek {\it et al.} \cite{B134434} have recently provided a key to this
long-standing problem by reanalyzing extensive observations of
antiferromagnetic rings.
When $1/T_1$ is divided by the static susceptibility-temperature product
$\chi T$, its peak is more pronounced and well-fitted to the
Lorentzian-type expression,
$\omega_{\rm c}(T)/[\omega_{\rm c}^2(T)+\omega_{\rm N}^2(H)]$, where
$\omega_{\rm N}\equiv\gamma_{\rm N}H$ is the Larmor frequency of probe
nuclei, whereas $\omega_{\rm c}$ is what they define as the
temperature-dependent correlation frequency.
The renormalized relaxation rate is thus peaked at a temperature
satisfying $\omega_{\rm c}(T)\simeq\omega_{\rm N}(H)$.
Considering the significant difference between the electronic and nuclear
energy scales ($\hbar\omega_{\rm N}\alt 10^{-5}|J|$),
$\hbar\omega_{\rm c}(T)$ may be ascribed to the averaged lifetime
broadening of discrete energy levels.
Demonstrating that chromic and ferric wheels give
$\omega_{\rm c}\propto T^\alpha$ with $\alpha=3.0\sim 3.5$,
Baek {\it et al.} claim that the exchange-coupled ion spins are likely to
interact with the host molecular crystal through the Debye-type phonons.

   In response to this stimulative report, several authors
\cite{S077203,S212402} inquired further into the underlying scenario.
While their arguments were elaborately based on microscopic spin
Hamiltonians and enlighteningly verified the relevance of spin-phonon
coupling to the notably peaked $1/T_1$, any magnetic anisotropy was
neglected and/or most of the transition matrix elements were discarded in
their evaluation of the dynamic spin correlation functions.
Thus, we take the naivest but thus cumbersome approach:
We set up realistically dressed Hamiltonians, completely diagonalize them,
and sum up all the transition matrix elements into the dynamic structure
factor.
Such calculations are inevitably restricted to sufficiently small clusters
but can nevertheless illuminate key factors in nanoscale spin
dynamics, {\it intrinsic intracluster anisotropies and extrinsic
intercluster phonons}.

   We consider both ferromagnetic and antiferromagnetic rings of various
metal ion spins and describe them by the Hamiltonian
\begin{eqnarray}
   &&\!\!\!\!\!\!\!\!\!\!\!\!\!
   {\cal H}
  =J\sum_{l=1}^L
   \mbox{\boldmath$S$}_{l}\cdot\mbox{\boldmath$S$}_{l+1}
  +\sum_{l=1}^L
   \mbox{\boldmath$d$}_{l}\cdot
   \bigl(\mbox{\boldmath$S$}_{l}\times\mbox{\boldmath$S$}_{l+1}\bigr)
   \nonumber\\
   &&\!\!\!\!\!\!\!\!\!\!\!
  +D_{\rm FM}\biggl(\sum_{l=1}^LS_l^z\biggr)^{\!\!2}
  +D_{\rm AFM}\sum_{l=1}^L\bigl(S_l^z\bigr)^{\!2}
  -g\mu_{\rm B}H\sum_{l=1}^LS_l^z,
   \label{E:H}
\end{eqnarray}
where $\mbox{\boldmath$S$}_{L+1}=\mbox{\boldmath$S$}_{1}$.
The first term is the isotropic Heisenberg exchange interaction.
The second term introduces the antisymmetric Dzyaloshinsky-Moriya (DM)
interaction with $\mbox{\boldmath$d$}_{l}$ set for $(-1)^l(d,0,0)$.
The third and fourth terms describe the axial magnetic anisotropy in the
cases of ferromagnetic ($J<0$) and antiferromagnetic ($J>0$) exchange
coupling, respectively.
Both $D_{\rm FM}$ and $D_{\rm AFM}$ are simply written as $D$ in all the
figures.
The last term represents the Zeeman interaction with an external field.
The nuclear spin-lattice relaxation rate is generally expressed as
\begin{equation}
   \frac{1}{T_1}
  =\frac{\gamma_{\rm N}^2}{2L}
   \sum_{\sigma\tau}\sum_q
   |A_q^{\sigma\tau}|^2 S^{\sigma\tau}(q,\omega_{\rm N}).
   \label{E:T1}
\end{equation}
$|A_q^{\sigma\tau}|^2$ is the form factor describing the
hyperfine coupling of spin excitations at wave vector $q$ with the
probe nuclei.
$S^{\sigma\tau}(q,\omega)$ is the dynamic spin structure factor and is
given by
\begin{eqnarray}
   &&
   S^{\sigma\tau}(q,\omega)
  =\sum_{i,j}{\rm e}^{-\hbar\omega_i/k_{\rm B}T}
   \langle i|S_q^\sigma|j\rangle\langle j|S_{-q}^\tau|i\rangle
   \nonumber \\
   &&\qquad\times
   \delta(\omega+\omega_i-\omega_j)
  /\sum_{i,j}{\rm e}^{-\hbar\omega_i/k_{\rm B}T},
   \label{E:Sqw}
\end{eqnarray}
where $S_q^\sigma$ is the Fourier transform of $S_l^\sigma$ and
$|i\rangle$ is an exact eigenstate of the Hamiltonian (\ref{E:H}) with
energy $\hbar\omega_i$.
There may be longitudinal contributions $S^{zz}(q,\omega_{\rm N})$ to
$1/T_1$ in general, originating from anisotropic hyperfine coupling, and
therefore strongly depending on the detailed molecular structure.
Considering that the characteristic peak of $1/T_1$ versus $T$ is commonly
observed for various ring nanomagnets, here, we assume that the transverse
spin fluctuations $S^{+-}(q,\omega_{\rm N})$ predominate in the
relaxation rate (\ref{E:T1}).
We further replace $A_q^{+-}$ by $A_{q=0}^{+-}\equiv A^{+-}$
\cite{L6946,L6839}, because the geometric coefficients $A_q^{\sigma\tau}$
usually exhibit little momentum dependence, particularly in proton NMR
measurements \cite{S077203,L3773}.
In the limit of $H\rightarrow 0$ without any anisotropy,
$S^{+-}(q,\omega)$ is equivalent to $S^{zz}(q,\omega)$ and can be
obtained rather easily.
However, it is not the case with any NMR measurement.
Even in the case of $d=0$, where the Hamiltonian (\ref{E:H}) commutes with
the total magnetization and a translation by one site along the ring, it
takes a long time to straightforwardly evaluate $S^{+-}(q,\omega)$ for
a hexanuclear Fe$^{\rm III}$ ring (Fe6) or an octanuclear Cr$^{\rm III}$
ring (Cr8).
With nonvanishing DM interaction, the naivest treatment of Fe6 and Cr8
with the Hamiltonian (\ref{E:H}) is impossible to begin with.
Therefore, we try to extract the essential physics from more tractable
clusters.
\begin{figure}
\centering
\includegraphics[width=83mm]{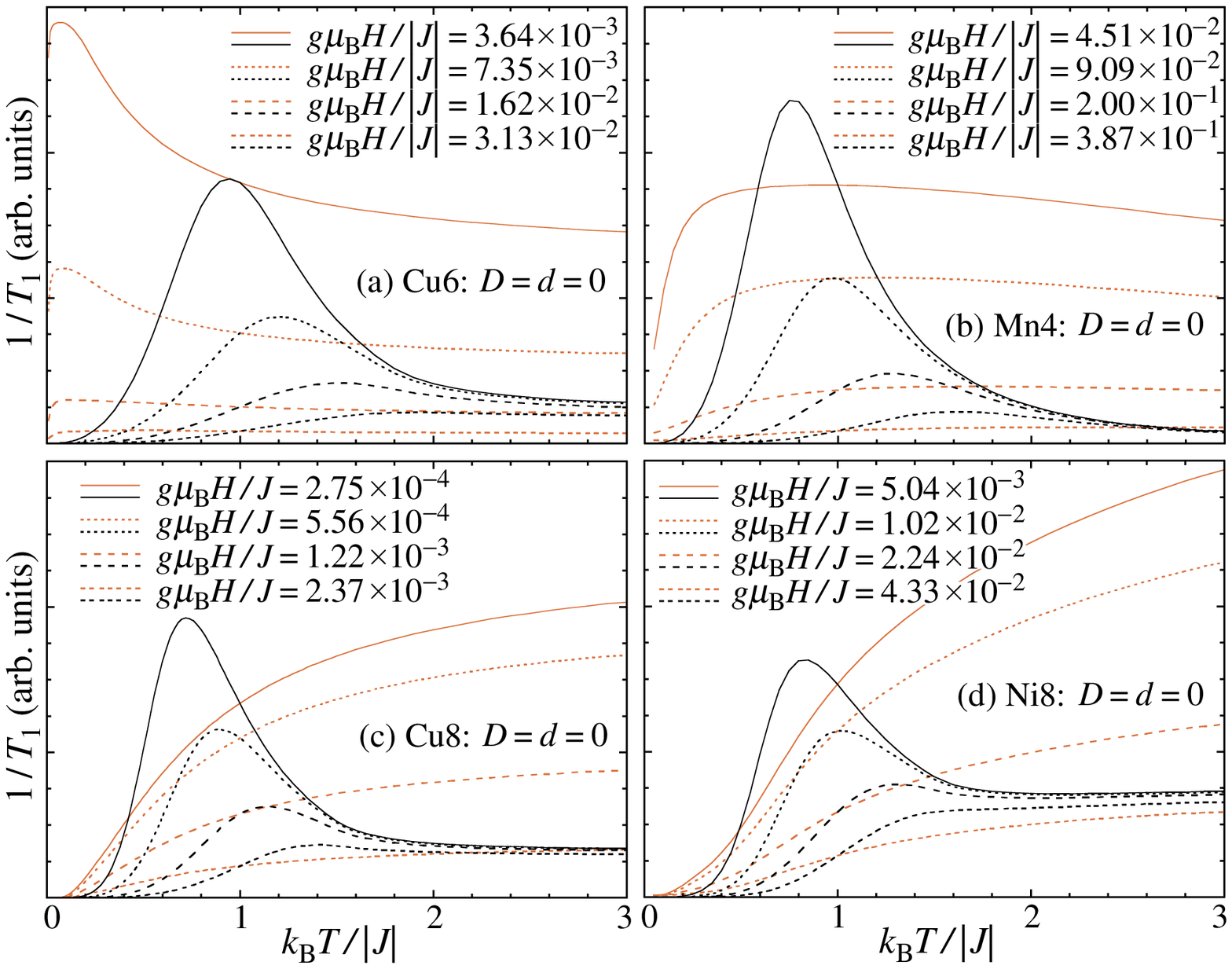}
\vspace*{-3mm}
\caption{(Color online)
         $1/T_1$ versus $T$ with varying fields applied to Cu6 (a),
         Mn4 (b), Cu8 (c), and Ni8 (d), which are assumed to be free from
         any anisotropy.
         Red (toned-down) and black lines show calculations on the
         assumption that ${\mit\Gamma}$ is independent of $T$ and
         proportional to $T^3$, respectively.}
\label{F:isotropic}
\vspace*{4mm}
\centering
\includegraphics[width=85mm]{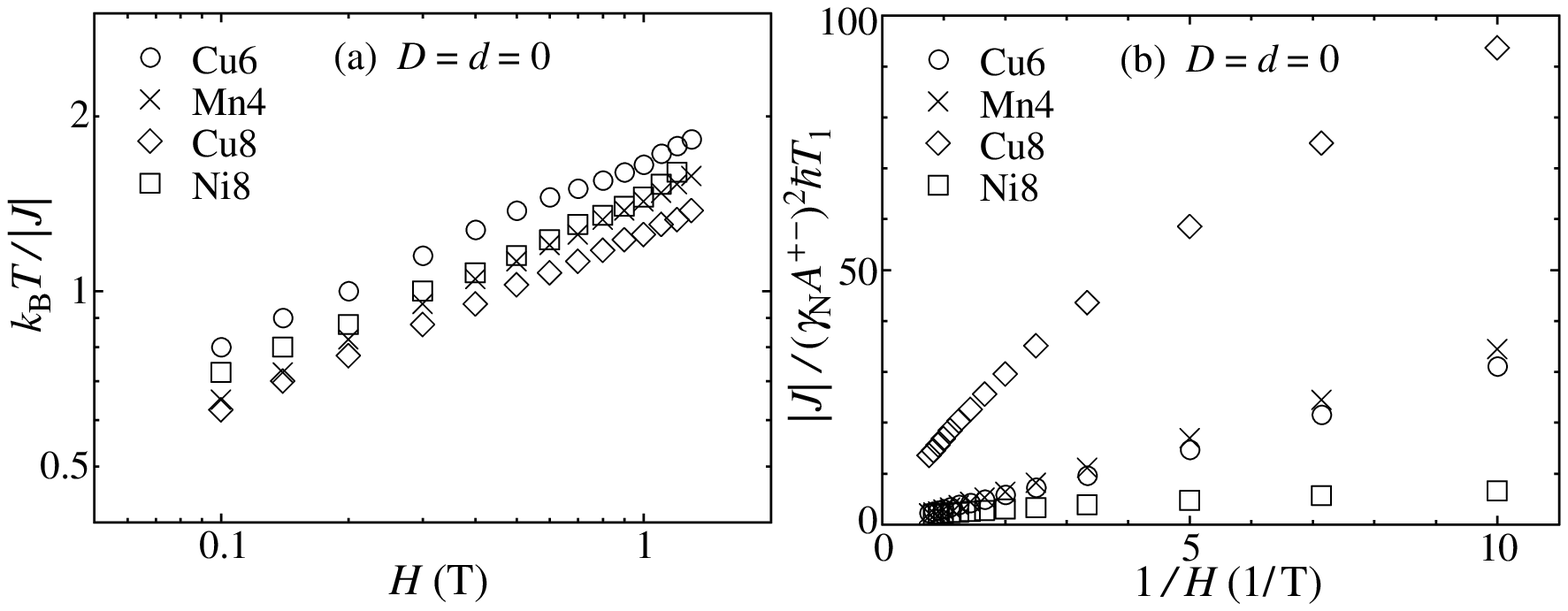}
\vspace*{-3mm}
\caption{(a) Temperature at which $1/T_1$ reaches its maximum is plotted
             versus $H$.
         (b) Maximum value of $1/T_1$ is plotted versus $1/H$,
             where any anisotropy is set equal to zero, and
             ${\mit\Gamma}$ is assumed to be proportional to $T^3$.}
\label{F:peak}
\end{figure}

\begin{figure*}
\centering
\includegraphics[width=176mm]{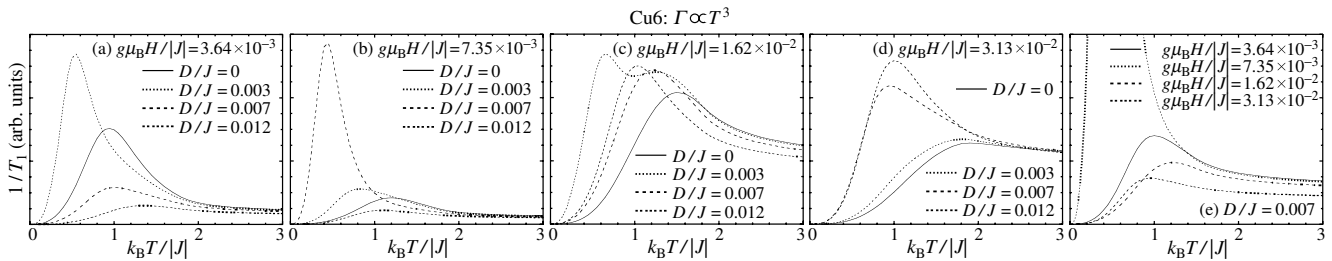}
\vspace*{-3mm}
\caption{$1/T_1$ versus $T$ for Cu6 with anisotropic exchange
         interaction,
         where ${\mit\Gamma}$ is set proportional to $T^3$.}
\label{F:Cu6}
\end{figure*}

\begin{figure*}
\centering
\includegraphics[width=176mm]{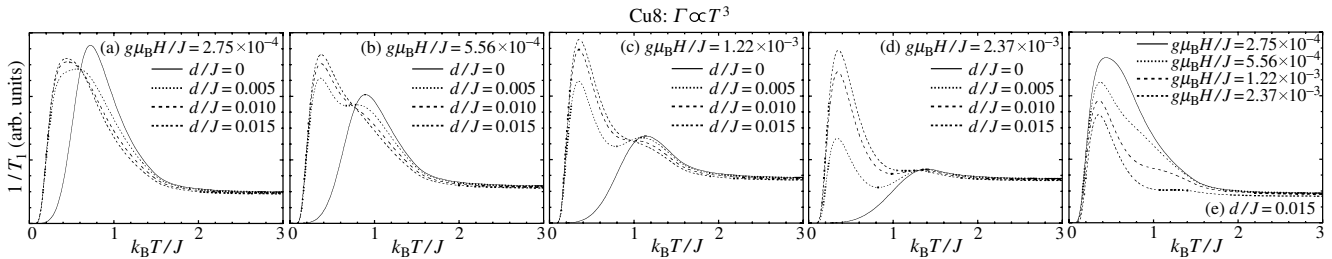}
\vspace*{-3mm}
\caption{$1/T_1$ versus $T$ for Cu8 with antisymmetric interaction,
         where ${\mit\Gamma}$ is set proportional to $T^3$.}
\label{F:Cu8}
\end{figure*}

   The most important ingredient in calculating eq. (\ref{E:Sqw}) is a
treatment for $\delta(\omega+\omega_i-\omega_j)$.
When we adopt the Lorentzian-type expression
\begin{equation}
   \delta(\omega+\omega_i-\omega_j)
   \simeq
   \frac{{\mit\Gamma}/\pi}{(\omega+\omega_i-\omega_j)^2+{\mit\Gamma}^2},
\end{equation}
$\hbar{\mit\Gamma}$ may be recognized as the lifetime broadening of the
discrete energy levels \cite{B134434}.
There has been a pioneering attempt \cite{S212402} to solve a coupled
spin-phonon Hamiltonian, leading to a level-dependent expression of
${\mit\Gamma}$.
However, such a calculation artificially selects particular transition
matrix elements out of the summation in eq. (\ref{E:Sqw}) and is feasible
in a rather limited situation, for instance, at sufficiently low
temperatures $k_{\rm B}T\ll\hbar\omega_{\rm D}$, where
$\hbar\omega_{\rm D}/k_{\rm B}$ is the Debye temperature, which is usually
less than $100\,\mbox{K}$ for molecular nanomagnets \cite{G2015}.
Here, we place great emphasis on including full spin degrees of freedom
into our calculation, and therefore discuss ${\mit\Gamma}$ in an empirical
way.
We set ${\mit\Gamma}$ proportional to $T^3$ on the one hand and independent
of temperature on the other hand.
The former treatment is motivated on the basis of the experimental
observations by Baek {\it et al.} \cite{B134434} and results in supporting
their scenario---discrete energy levels of antiferromagnetic rings are
lifetime-broadened owing to spin-acoustic phonon coupling.
The latter treatment attempts to exclude any possibility of phonons
assisting the nuclear spin-lattice relaxation and results in suggesting
more intrinsic mechanism such as intercluster dipolar interactions
\cite{L224401} for the lifetime broadening of discrete energy levels.

   First, we consider various magnetic rings without any anisotropy in an
attempt to elucidate the effects of acoustic phonons on the relaxation
rate.
Figure \ref{F:isotropic} shows the temperature dependences of $1/T_1$ for
ferromagnetic hexanuclear Cu$^{\rm II}$ (Cu6) \cite{R4427},
ferromagnetic tetranuclear Mn$^{\rm III}$ (Mn4) \cite{B14046},
antiferromagnetic octanuclear Cu$^{\rm II}$ (Cu8) \cite{A4347}, and
antiferromagnetic octanuclear Ni$^{\rm II}$ (Ni8) \cite{X2950} rings.
When we set ${\mit\Gamma}$ to be independent of $T$, as well as $H$, the
resultant $1/T_1$ is reminiscent of $\chi T$; that is to say, it
monotonically decreases and increases with $T$ in the ferromagnetic and
antiferromagnetic cases, respectively.
No peak appears in the curve of $1/T_1$ versus $T$ except for a trivial
ceiling due to the Zeeman splitting of the ferromagnetic ground state.
If we adopt ${\mit\Gamma}$ to be proportional to $T^3$, a notable peak
appears at $T=(T)_{\rm peak}\simeq |J|/k_{\rm B}$, which is
qualitatively consistent with numerous experimental observations
\cite{L14341,L1115,L6946,B134434,P184419,M034710,S077203}.
We have confirmed that such a peak structure does not appear with any
other temperature dependence of ${\mit\Gamma}$, whether it be of power-law
type or of activated type.
With increasing field, the peak becomes smaller and shifts to higher
temperatures monotonically.
Figure \ref{F:peak}(a) shows $(T)_{\rm peak}$ versus $H$ and reveals its
power-law behavior $(T)_{\rm peak}\propto H^\nu$.
We find that $\nu\simeq 0.32$, $0.34$, $0.30$ and $0.31$ for Cu6, Mn4, Cu8
and Ni8, respectively.
Considering the significant difference between the nuclear and electronic
energy scales, $\hbar\omega_{\rm N}\sim 10^{-5}J$, it must be the averaged
lifetime broadening of discretized energy levels, referred to as
$\hbar\omega_{\rm c}$ \cite{B134434}, that characterizes $1/T_1$.
The relaxation should be most accelerated at
$\omega_{\rm N}=\gamma_{\rm N}H\simeq\omega_{\rm c}$; that is, we suppose
that $1/T_1$ is proportional to
$\omega_{\rm c}/(\omega_{\rm c}^2+\omega_{\rm N}^2)$.
Then, the above findings indicate that $\omega_{\rm c}\propto T^\alpha$,
with $\alpha$ ranging from $2.9$ to $3.3$, which is in very good agreement
with some experimental findings \cite{B134434}.
However, there are observations of $(T)_{\rm peak}$ remaining almost
unchanged with $H$ \cite{L14341,L1115}.
It is also interesting to plot the maximum value of $1/T_1$,
$(1/T_1)_{\rm max}$, versus $1/H$.
Figure \ref{F:peak}(b) shows that in the ferromagnetic cases, there well
holds a linear relation between them, which is again consistent with the
above consideration,
$1/T_1\propto\omega_{\rm c}/(\omega_{\rm c}^2+\omega_{\rm N}^2)
 \leq 1/2\gamma_{\rm N}H$, whereas in the antiferromagnetic cases, the
calculations deviate from such a straightforward relationship.
The peak structure indeed depends not only on the constituent metal ion
spins, but also on their surrounding ligands \cite{L14341}.
We are thus led to consider the effects of magnetic anisotropy.
${\mit\Gamma}$ is always set proportional to $T^3$ in the following.

   The ferromagnetic ring Cu6 is simulated in more detail in
Fig. \ref{F:Cu6}, where the anisotropic exchange interaction $D_{\rm FM}$
is taken into account.
The anisotropy effect is so significant at $k_{\rm B}T\simeq |J|$
[Figs. \ref{F:Cu6}(a)$-$\ref{F:Cu6}(d)] as to completely break the
monotonic field dependences of both $(T)_{\rm peak}$ and
$(1/T_1)_{\rm max}$ [Fig. \ref{F:Cu6}(e)].
Even a double peak may appear in $1/T_1$ versus $T$.
Metallic wheels \cite{C1379,A6443,S2482,A1430} can act as hosts for
an alkali-metal ion, which affects the coordination geometry of the
environmental transition-metal ions.
The experimental findings for hexagonal ferric wheels with a lithium ion,
a sodium ion and no alkali ion at the center are indeed different from
each other in a subtle way \cite{L14341,P184419} and such observations are
consistent with the sensitivity of $1/T_1$ to magnetic anisotropy.

   The antiferromagnetic ring Cu8 is free from the crystalline anisotropy
$D_{\rm AFM}$.
On the other hand, $^{63}$Cu nuclear quadrupole resonance spectra of Cu8
rings indicate four crystallographically inequivalent Cu$^{\rm II}$ ions
in each ring, which is attributable to the slight deviation of
Cu$^{\rm II}$ ions from a planar octagon \cite{A4347} and/or possible
distortion of the octagonal environment due to spin-lattice coupling
\cite{S197202}.
Hence, we introduce a DM term into the Hamiltonian of Cu8 and show the
resultant $1/T_1$ in Fig. \ref{F:Cu8}.
The DM interaction induces another peak at a temperature much lower than
$J/k_{\rm B}$ [Figs. \ref{F:Cu8}(a)$-$\ref{F:Cu8}(d)].
As soon as $d$ is switched on, there appears a renewed peak at
$k_BT\simeq 0.4J$.
With increasing $d$, the peak remains sitting at almost the same
temperature, whereas its height monotonically increases and reaches
saturation.
Thus, a DM term in the Hamiltonian masks the peak structure at
$k_{\rm B}T\simeq J$.
Both $(T)_{\rm peak}$ and $(1/T_1)_{\rm max}$ for the renewed peak
monotonically decrease with increasing field, but their field dependences
are much more moderate than those of the original peak
[Fig. \ref{F:Cu8}(e)].

\begin{figure}
\centering
\includegraphics[width=86mm]{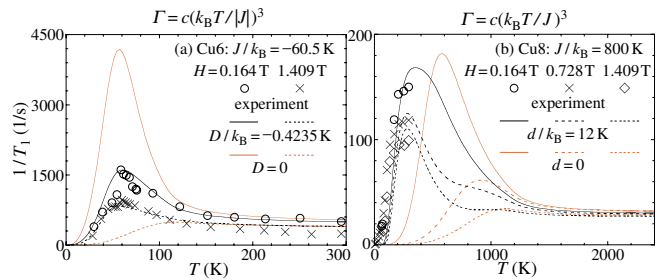}
\vspace*{-6mm}
\caption{(Color online)
         $1/T_1$ versus $T$ for proton nuclei in
         $[(\mbox{PhSiO}_2)_6\mbox{Cu}_6(\mbox{O}_2\mbox{SiPh})_6]\cdot
          6\mbox{EtOH}$ (a) and
         $[\mbox{Cu}_8(\mbox{C}_5\mbox{H}_7\mbox{N}_2)_8(\mbox{OH})_8]$
         (b).
         Experimental findings \cite{L6946,L6839} (symbols) are compared
         with calculations with (black lines) and without [red (toned-down)
         lines] anisotropies, where ${\mit\Gamma}$ is set
         proportional to $T^3$.}
\label{F:comparison}
\end{figure}

   Finally, we attempt to reproduce experimental findings for the
ferromagnetic hexanuclear Cu$^{\rm II}$ rings
$[(\mbox{PhSiO}_2)_6\mbox{Cu}_6(\mbox{O}_2\mbox{SiPh})_6]\cdot
 6\mbox{EtOH}$ \cite{R4427}
and the antiferromagnetic octanuclear Cu$^{\rm II}$ rings
$[\mbox{Cu}_8(\mbox{C}_5\mbox{H}_7\mbox{N}_2)_8(\mbox{OH})_8]$
\cite{A4347} within the framework of our theory.
We consider an anisotropic exchange contribution $D/|J|=0.007$ with
$J/k_{\rm B}=-60.5\,\mbox{K}$ for the Cu6 clusters \cite{R4427}, while
alternating DM vectors $d/J=0.015$ with $J/k_{\rm B}=800\,\mbox{K}$ for
the Cu8 clusters \cite{L6946}.
Figures \ref{F:comparison}(a) and \ref{F:comparison}(b) show the
thus-calculated $1/T_1$ for the Cu6 and Cu8 clusters, respectively, where
the lifetime broadening of discrete energy levels is taken as
${\mit\Gamma}=c(k_{\rm B}T/|J|)^3$ with
$c=3.2\times 10^{10}\,\mbox{rad}\cdot\mbox{Hz}$ (Cu6) and
$c=1.0\times 10^{11}\,\mbox{rad}\cdot\mbox{Hz}$ (Cu8), while
the hyperfine coupling strength is set for
$(\gamma_{\rm N}A^{+-})^2=18\times 10^{14}\,\mbox{rad}^2\mbox{Hz}^2$
(Cu6) and
$(\gamma_{\rm N}A^{+-})^2=2.8\times 10^{14}\,\mbox{rad}^2\mbox{Hz}^2$
(Cu8).
The present coupling constants are so reasonable as to fit with
the experimental estimates \cite{L6946}:
$(\gamma_{\rm N}A^{+-})^2=(8\pm 2)\times 10^{14}\,\mbox{rad}^2\mbox{Hz}^2$
($^1$H NMR on Cu6) and
$(\gamma_{\rm N}A^{+-})^2=(9\pm3)\times 10^{14}\,\mbox{rad}^2\mbox{Hz}^2$
($^1$H NMR on Cu8)
within a factor of about three.
In the Cu6 clusters, $1/T_1$ is sensitive to the anisotropic exchange
interaction at $k_{\rm B}T\alt 2|J|$ in particular, and that is why we find
such a reduced field dependence of $(1/T_1)_{\rm max}$ in
Fig. \ref{F:comparison}(a).
The ratio of
$(1/T_1)_{\rm max}(H=0.164\,\mbox{T})$ to
$(1/T_1)_{\rm max}(H=1.409\,\mbox{T})$ is two or less, whereas in another
measurement on the same material \cite{L1115} the ratio is reported to
reach three.
Considering that both measurements were performed on polycrystalline
powders, such a sample dependence of $T_1$ findings sounds convincing.
There exists a single crystal of Mn4 ferromagnetic rings instead
\cite{B14046}, where any anisotropy effect can be verified quantitatively.
In the Cu8 clusters, the DM interaction appears to be essential for peaks
to appear at much lower temperatures than $J/k_{\rm B}$.
The measurement is unfortunately limited to $k_{\rm B}T\alt 0.5J$, because
the Cu8 sample breaks down at above room temperature.
The observations nevertheless suggest that $1/T_1$ almost reaches
saturation at room temperature.
The probable low-temperature peaks in the Cu8 clusters may be
characteristic of an antisymmetric interaction and should be
distinguished from peaks originating from the exchange interactions,
whether isotropic or anisotropic, which usually appear at
$k_{\rm B}T\simeq J$.

   We have microscopically interpreted the temperature dependences of
$1/T_1$ in molecular ring nanomagnets.
In comparison with antiferromagnetic rings
\cite{S212402,S197202,S014418,W024424,A}, ferromagnetic ones have so far
much less been calculated microscopically.
Many of antiferromagnetic features were derived from isotropic
Hamiltonians and were hardly compared with corresponding experimental
observations.
In such circumstances, we have investigated both ferromagnetic and
antiferromagnetic rings with particular emphasis on possible anisotropic
terms in their Hamiltonians and their full diagonalization.
While $1/T_1$ versus $T$ in exchange-coupled ion spins of ring shape may
be notably peaked at $k_{\rm B}T\simeq |J|$ provided they are coupled to
acoustic phonons of the host molecular crystal, the peak structure is
very sensitive to magnetic anisotropies.
With increasing anisotropies due to single-ion and dipolar interactions,
which are all describable within symmetric spin-bilinear terms, both
$(1/T_1)_{\rm max}$ and $(T)_{\rm peak}$ move in an irregular way.
With increasing antisymmetric DM interaction, on the other hand, another
peak grows monotonically far from the original peak at
$k_{\rm B}T\simeq J$.
The sample-dependent field dependence of $1/T_1$ for Cu6 clusters
\cite{L1115,L6946} and the almost saturated $1/T_1$ at room temperature
for Cu8 clusters \cite{L6946,L6839} are both understandable in this
context.
Indeed anisotropic contributions to the Hamiltonian are no more than a few
percent of the exchange coupling constant in most cases
\cite{R4427,C12177,C461,C094405}, but they potentially cause a drastic
change in $1/T_1$ particularly at low temperatures $k_{\rm B}T\alt |J|$.
With increasing field, the $1/T_1$ peak becomes smaller and shifts to
higher temperatures in Cr8 antiferromagnetic rings \cite{B134434}, which
is reminiscent of Figs. \ref{F:isotropic}(c) and \ref{F:isotropic}(d),
while the peak remains sitting at almost the same temperature in Fe6
antiferromagnetic rings \cite{L14341,L1115}, which may be ascribed to
alternating DM vectors \cite{P184419,C461}.
With nonvanishing DM vectors, neither Cr8 nor Fe6 clusters can be fully
diagonalized.
Low-temperature observations, such as quantum fluctuations of the
ground-state spin multiplet \cite{Y157603} and level crossings in an
applied field \cite{C88}, may be understandable within an effective theory
based on selected low-lying states, whereas the essentially thermal
behavior of our interest should be examined with the whole Hamiltonian.
Understanding will increase with further measurements of
$[\mbox{Mn}_4\mbox{Cl}_4(\mbox{C}_9\mbox{H}_9\mbox{NO}_2)_4]$
\cite{B14046} and
$[(\mbox{C}_{24}\mbox{H}_{16}\mbox{N}_8\mbox{O}_2)_4\mbox{Ni}_8
  (\mbox{H}_2\mbox{O})_8](\mbox{BF}_4)_8\cdot16\mbox{H}_2\mbox{O}$
\cite{X2950}, which are ideal ferromagnetic and antiferromagnetic rings,
respectively, with relatively small spin degrees of freedom and moderate
intramolecular exchange interactions.

   We thank S. Miyashita, Y. Furukawa, F. Borsa, and S. Maegawa for
fruitful discussions.
We give our warmest acknowledgments to Y. Furukawa and F. Borsa for
providing their experimental data.
This study was supported by the Ministry of Education, Culture, Sports,
Science, and Technology of Japan.

\end{document}